\documentclass[11pt]{article}
\usepackage[margin=1.05in]{geometry}
\usepackage[T1]{fontenc}
\usepackage[utf8]{inputenc}
\usepackage{booktabs,array,microtype,amsmath,url,graphicx}
\usepackage[hidelinks]{hyperref}
\usepackage[font=small,labelfont=bf]{caption}
\usepackage{listings}
\newcommand{\code}[1]{\texttt{\small #1}}

\title{\bfseries What Stops a Small Language Model\\ From Driving a Database Agent}
\author{
  Cevheri Bozo\u{g}lan \and Yusuf G\"{u}ndo\u{g}du \and Abdullah Kaya \and Koray \c{S}irin\\[4pt]
  \normalsize Independent researchers, and contributors to the software evaluated\\
  \normalsize \texttt{github.com/libredb/libredb-studio}
}
\date{\today}

\begin{document}
\maketitle

\begin{abstract}
\noindent
Small open-weight language models are assumed to fail at agentic database work because they lack the reasoning
capacity for it. We test that against a production system. Over eleven days we drove the agent mode of an
open-source SQL client with 39 open-weight models served locally and one hosted control, across six task
surfaces: 8{,}199 runs, 110{,}711 ledger events, 14{,}008 refused tool calls. Of the 2{,}100 model-attributed
agent-mode losses, 1{,}590, or 75.7\%, came from runs that had invoked at least one tool. That majority is what
survives resampling models rather than runs: it holds in 99.7\% of clustered resamples and in 15 of the 22 models
with at least twenty losses. Within it, transport, a run that used the tools and never got a deliverable through,
is the largest class at 36.2\% and capability, a run that invoked no tool at all, the smallest at 17.3\%; we
report that ordering as a property of this corpus rather than a general finding, since clustered by model it
holds in only 74.5\% of resamples. Transport failures decompose into a few mechanical argument shapes. Production
ledgers record refusal codes and never the model's arguments, so these were invisible for ten days; capturing
them exposed five server defects, one of which demanded a field on one tool, forbade it on the sibling that
composed it, then failed the run for its absence. Five server changes, touching no model, prompt or sampling
setting, moved six models by 6 to 21 cells out of 30. We also report a confound we believe affects published
local-model benchmarks, ours included: with no context cap, one 7.1 GB model was admitted at its full
262{,}144-token window and held 51 GB on a 64 GB machine, producing runs indistinguishable in any ordinary log
from a model timing out. The corpus, the scorer and a verifier that regenerates every figure are released.
\end{abstract}

\section{Introduction}

The question we set out to answer was operational. LibreDB Studio ships an agent mode that plans and executes
read-only database work through a fixed tool contract. Its users run it against their own databases, and a
substantial fraction want to run it against a model on their own hardware rather than a hosted API, because
sending schema inventories and query text to a third party is exactly what their environment forbids. Which
locally hosted models can actually drive that agent, and what has to change so that more of them can?

The prevailing framing, in vendor documentation and practitioner writing alike, is a capability threshold: below
some parameter count a model cannot sustain a tool-calling loop. That framing makes a specific, checkable
prediction. Failures should look like models that do not act. What we observe is close to the opposite. Most
failures are models that act, establish something true, and then lose it at the boundary between the model and
the server.

We are not the first to report that agent harnesses are implicated in what looks like model failure, and we say
so at the outset. Section~\ref{sec:related} places this work against results that predate it. Our contribution
is a field record at production scale, with the ledger, the extractor and the argument captures released, and
with our own methodological choices and confounds reported rather than smoothed.

This paper contributes:

\begin{enumerate}
\item A four-class failure taxonomy for agentic runs, capability, transport, clock and verification, computable
over a structured run ledger, together with its measured distribution across 40 models and six surfaces, and an
explicit sensitivity analysis of the one classification decision that changes the headline.
\item A characterisation of the transport class from captured tool-call arguments, showing that it decomposes
into a small number of mechanical shapes in which the model was substantively right.
\item Five server-side interventions with before and after readings on the same hardware, prompts and tasks,
including an honest separation of what did and did not move the numbers.
\item Two measurement confounds, unbounded serving context and swap pressure, that produce results
indistinguishable from model failure and that we believe affect published local-model evaluations.
\end{enumerate}

\section{Related work}
\label{sec:related}

\paragraph{Feedback quality as the binding constraint.}
Olausson et al.~\cite{olausson2024selfrepair} showed that self-repair in code generation is bottlenecked by the
quality of the feedback rather than by the repairer's ability to act on it. Closest to the present work,
Gumaan~\cite{gumaan2026feedback} measured the change in log-probability of re-emitting an action after its
failure is recorded in the transcript, found this quantity negative for every instruction-tuned model tested,
six checkpoints from 135M to 1.7B across four families, and reported the probability of repeating a failed call
rising from 0.06 to 0.54. Counterfactuals there attribute 83\% of the damage to the failed call's surface form
in the context rather than to the semantics of marking it failed, and show that replacing the verbatim call with
a runtime-generated description of the failure removes 76\% of the repetition. That work establishes both the
phenomenon and a direction of repair, and it predates our measurements. Wang et al.~\cite{wang2024mint}
quantified feedback's contribution in multi-turn interaction and found that better single-turn performance does
not imply better multi-turn performance. Huang et al.~\cite{huang2024selfcorrect} ruled out intrinsic
self-correction as the mechanism, and Gou et al.~\cite{gou2024critic} showed that external tool feedback is what
drives correction.

We differ in three ways. Our models are an order of magnitude larger, 1.7B to 35B for the tags that name a size, rather than 135M to 1.7B. Our
outcome is end-to-end task success under a server-side verifier rather than a log-probability proxy. And our
setting is a deployed product against a live database, so the evidence check runs inside the loop: a report is
refused by the server during the run rather than scored after it, which is what makes verification a cause of
loss here rather than a label applied to a finished trace.

\paragraph{The same failure, repaired by training.}
Zhang et al.~\cite{zhang2026fission} report independently that smaller models after an execution error often
fall into repetitive invalid re-invocations, and address it with reinforcement learning on simulated diagnostic
feedback. Vuddanti et al.~\cite{vuddanti2025paladin} train on failure-injected trajectories. Kadekodi et
al.~\cite{kadekodi2025agentflux} decouple tool selection from argument generation and find argument generation
the weaker half, which is exactly where our refusals occur. Our interventions require no gradient updates and
help every model at once, which is the practical difference.

\paragraph{Tool-use benchmarks.}
BFCL~\cite{patil2025bfcl}, ToolLLM~\cite{qin2024toolllm}, API-Bank~\cite{li2023apibank},
ToolSandbox~\cite{lu2024toolsandbox} and $\tau$-bench~\cite{yao2024taubench} vary the model against a harness
whose rejection messages are fixed and form part of the benchmark. Holding those messages fixed is what makes
scores comparable across models, and it is also what places the harness outside what those five can measure.

\paragraph{The harness as an experimental variable.}
That axis is not ours to claim. Sigdel and Baral isolate the tool interface
directly~\cite{sigdel2026schemafirst}, comparing free-form documentation, JSON Schema, and JSON Schema with
structured validation diagnostics under identical tool semantics, and report that the schema conditions reduce
interface misuse but not semantic misuse; that pilot runs one open local model and ends at zero task success in
every condition. Their ToolMisuseBench~\cite{sigdel2026toolmisuse} makes the interface a declared condition
rather than a fixture, with per-task fault plans replayed under a fixed seed across schema drift, rate limits,
timeouts, authorization failures and adversarial error rewriting, so the text returned after a failed call is
itself a variable; its baselines are deterministic policies rather than language models, and its environments are
simulators that need no external service. AgentCheck~\cite{mazumder2026agentcheck} records an agent's real tool
responses, replays them with exactly one perturbed, and re-runs under a mitigation wrapper against the identical
fault, which is a reproduce-and-intervene loop like the one in Section~\ref{sec:interventions}, applied at the
tool layer rather than to the server's contract. Xiong et al.~\cite{xiong2025butterfly} hold the agent scaffold
fixed and apply fifteen perturbations to the three input sources a tool agent reads, the tool document, the user
query and the tool return, and close by recommending that tool error messages be redesigned so that models can
correct failures.

What this work adds is the setting and the provenance rather than the axis. The harness defects reported here
were not injected. They were live in a shipped validator, recovered from a production ledger, and then repaired
in the product, and the effect of repairing them is measured across 40 models rather than a scripted policy or a
single pilot model. The change those studies recommend is the change we made and then measured.

\paragraph{The same separation, the opposite ordering.}
ToolFailBench~\cite{soni2026toolfailbench} draws the distinction Table~\ref{tab:taxonomy} draws, between a run
that never called a tool and one that called it and then failed to use the result, and reports the first as much
the larger class across nineteen models. Its protocol is single-turn against mock tools that return a controlled
value, with the final answer written in a second step and no further turn in which to repair a rejected call, so
a call the system did not execute is recorded as a skip rather than answered with a refusal the model could act
on. Its own rule counts a tool call emitted as plain text in the answer body and never executed within that skip
class; we count the same observable as a transport failure of the surface that did not parse it, and repair it in
Section~\ref{sec:interventions}. We read the two results as the same argument from opposite ends. Which class a
tool-calling failure falls into is a joint property of the model and the harness, and a design that varies only
one of them settles the attribution in advance.

\paragraph{Agent benchmarks that attribute failure to capability.}
AgentBench~\cite{liu2024agentbench} evaluates across eight environments including a database environment and
attributes the gap between open and commercial models to long-term reasoning, decision making and instruction
following. It is both the nearest prior database-agent evaluation and the clearest statement of the position our
results complicate.

\paragraph{Database-side work.}
The text-to-SQL lineage, Spider~\cite{yu2018spider} and BIRD~\cite{li2023bird}, evaluates a single emitted
statement. SParC~\cite{yu2019sparc} and CoSQL~\cite{yu2019cosql} are multi-turn, but the counterpart in the loop
is a human, so they cannot exhibit this failure mode. Spider 2.0~\cite{lei2025spider2} is the nearest prior art
on the agent axis. D-Bot~\cite{zhou2024dbot} is the closest sibling to our setting, an agent that reads a live
system and writes a report. CodeS~\cite{li2024codes} makes small models competitive at text-to-SQL by training
them for it; our result concerns off-the-shelf models once the harness is repaired. Belcak et
al.~\cite{belcak2025slm} argue on general grounds that small language models suit agentic work, a position for
which this paper supplies field evidence without testing it directly.

\section{System under test}

LibreDB Studio's agent is a constrained loop with a verifier, not a general assistant with database tools
attached. Three properties matter for reading the results.

\paragraph{Six surfaces.} Five run in agent mode, \code{investigation}, \code{query-optimization},
\code{database-assessment}, \code{operations} and \code{data-analysis}, and one in planning mode. Each carries
its own objective, tool set and verifier. They are not difficulty tiers but different shapes of work. Planning is
toolless: the server hands a planning run an empty tool set, so no planning run can invoke a tool. The server
still reads the database before a planning run's first turn, with statements it composes itself; what planning
does not do is let the model choose a read. Data analysis must both report and \emph{present} a result. Query optimization must
produce a recommendation the user's editor can apply.

\paragraph{A tool contract with evidence.} Every claim in a report must cite an artifact the run actually
produced: the correlation id of a completed read, or the fingerprint of a schema snapshot the run captured.
Citations are checked against the run's own ledger and refused when they do not resolve. This is a deliberate
anti-hallucination measure and, as Section~\ref{sec:results} shows, a substantial source of loss.

\paragraph{A per-model settings layer.}
The server carries a settings layer that can override sampling, the per-call timeout, reminder limits and several
loop behaviours for a named model. Ten of the models in this corpus carry an entry in it, covering 1{,}013 of the
4{,}951 model-attributed runs; the apparatus in Section~4.2 is the default that applies to the rest. None of the
six models in the intervention table carries an entry: all six ran at the compiled defaults both before and
after, so no per-model setting differs between their two readings.

\paragraph{A verifier, not a human judge.} Each run ends with a verdict recording \code{answered} or
\code{unanswered} together with a list naming what was missing. The pass label is produced by the same code path
for every run, with no post-hoc scoring, which is what makes the corpus analysable at all. Critically,
\code{status: succeeded} is not a pass: a run that reports nothing also exits successfully. In our corpus 2{,}204
runs ended \code{model-stopped} and 1{,}405 of those were unanswered, so conflating process exit with task
success would inflate the apparent pass rate by that margin.

\section{Method}

\subsection{The cell and the lock}

The unit of measurement is a \emph{cell}: one model on one surface, run five consecutive times against a fixed
objective. Five consecutive passes locks the cell, and a model is complete at 30/30 over six cells. We report the
pair rather than a percentage, because the denominator is what makes two readings comparable.

Two bookkeeping rules govern this, and both were learned by violating them. A lock is never taken back: a cell at
5/5 is not re-measured or re-tuned. And a row must be read in one sitting under one configuration. Two models
were recorded at 30/30 and were not: their cells were genuinely 5/5, thirty runs and thirty passes all on the
ledger, but taken across three days and several settings. Read together in one sitting they returned 22/30 and
23/30. A model ships with \emph{a} reading, never with the union of readings.

\subsection{Apparatus}

\begin{table}[htbp]
\centering\small
\begin{tabular}{@{}l p{11.4cm}@{}}
\toprule
Machine & MacBook Pro (Mac17,6), Apple M5 Max, 18 cores, 64\,GB unified memory \\
Operating system & macOS 26.6 \\
Local inference & Ollama 0.33.0; context length unset for every figure in \S\ref{sec:results} (model default,
up to 262{,}144 tokens), capped at 32{,}768 only for the confirmations in \S\ref{sec:memory} \\
Runtime & Bun 1.4.2, Next.js 16 \\
Database & embedded SQLite sample, employee/department/salary schema \\
Turn limit & 90{,}000\,ms default; 150{,}000\,ms is the maximum any shipped model carries \\
Hosted control & \code{gemini-3.5-flash-lite} \\
\bottomrule
\end{tabular}
\caption{Apparatus. The sample database is deliberately small and fixed: we measure the model's ability to drive
the loop, not the database's ability to be large, and a fixed schema keeps the objective identical across all
8{,}199 runs.}
\end{table}

\subsection{What the ledger records, and what it does not}

Every run writes a framed-JSON event stream: run started, driver resolved, context captured, tool invoked, tool
completed, tool refused, call declined, call held, guidance issued, model stopped saying, report composed, answer
composed, run finished.

Analysis is done from the ledger and never from the sweep logs, and the distinction is not pedantic. A sweep log
records what a runner asked for; the ledger records what happened. They have disagreed materially: one chain
reported forty-eight completed runs in a minute having run none, and a log header printing \code{LLM\_PROVIDER=ollama}
belonged to runs the ledger shows went to a hosted API.

One omission in the ledger is deliberate and became the central methodological finding of this work. A declined
call records the tool, the refusal code and the validator's field paths, and never the model's arguments, so that
model-authored text cannot enter the server's own audit vocabulary. That is correct for a production ledger. It
also meant that for ten measurement days we could see \emph{that} a required field was absent and never
\emph{where the model had put it instead}. Section~\ref{sec:transport} reports what happened when we added a
temporary, environment-gated argument dump.

\subsection{Classification, and the decision that matters}
\label{sec:classdef}

We classify each loss by what the ledger shows the run did:

\begin{itemize}
\item \textbf{clock}: the run ended \code{model-timeout}, \code{turn-limit} or \code{deadline-exceeded}.
\item \textbf{capability}: the run invoked no tool at all.
\item \textbf{verification}: the run ended \code{report-composed} and was still unanswered, that is, a report was
filed and rejected.
\item \textbf{transport}: the run invoked tools, did not run out of time, and never got a deliverable through.
\end{itemize}

These are applied in that order, and \textbf{the order is load-bearing}. Of the 2{,}100 model-attributed agent-mode losses,
231 both ran out of clock and invoked no tool, so they satisfy the first two definitions simultaneously. We
report both assignments in Section~\ref{sec:results} and adopt clock-first, for a reason that is empirical rather
than stylistic: 225 of those 231 runs also emitted no text at all. A run that produced neither a tool call nor a
single character within its turn limit does not look like a model declining to act; it looks like a run that
never got going, which is the signature of the memory confound in \S\ref{sec:memory}. We regard the six runs that
did emit text, a median of about 2{,}000 characters, as genuinely mixed cases.

\section{Results}
\label{sec:results}

\subsection{Corpus}

\begin{table}[htbp]
\centering\small
\begin{tabular}{@{}lr@{}}
\toprule
Runs & 8{,}199 \\
Named models & 40 \\
\quad open-weight, served locally & 39 \\
\quad hosted, proprietary & 1 \\
Runs attributable to a named model & 4{,}951 \\
Runs predating the driver-resolved ledger event & 3{,}248 \\
Surfaces & 6 \\
Code checkouts the corpus spans & 5 \\
Answered & 4{,}983 (60.8\%) \\
Unanswered & 3{,}142 (38.3\%) \\
Runs with no recorded verdict & 74 (0.9\%) \\
Refused tool calls & 14{,}008 \\
\bottomrule
\end{tabular}
\caption{The corpus. Run counts per named model range from 3 to 629, with 35 models at 20 runs or more.}
\end{table}

\subsection{Pass rate by surface}

\begin{table}[htbp]
\centering\small
\begin{tabular}{@{}lrrr@{}}
\toprule
Surface & Answered & $n$ & Rate \\
\midrule
\code{planning} & 466 & 560 & 83.2\% \\
\code{investigation} & 389 & 550 & 70.7\% \\
\code{operations} & 404 & 612 & 66.0\% \\
\code{database-assessment} & 342 & 614 & 55.7\% \\
\code{query-optimization} & 639 & 1{,}360 & 47.0\% \\
\code{data-analysis} & 468 & 1{,}206 & 38.8\% \\
\bottomrule
\end{tabular}
\caption{Pass rate by surface, model-attributed runs only. The ordering does not track the intuitive difficulty
of the question. It tracks how many schema-constrained objects the model must emit: planning asks for one
statement, data analysis asks for a presented answer and a cited report.}
\end{table}

\subsection{The failure taxonomy}

\begin{table}[htbp]
\centering\small
\begin{tabular}{@{}lrrrr@{}}
\toprule
& \multicolumn{2}{c}{clock-first (adopted)} & \multicolumn{2}{c}{capability-first} \\
\cmidrule(lr){2-3}\cmidrule(lr){4-5}
Class & Count & Share & Count & Share \\
\midrule
transport & 761 & \textbf{36.2\%} & 761 & 36.2\% \\
clock & 542 & 25.8\% & 395 & 18.8\% \\
verification & 434 & 20.7\% & 434 & 20.7\% \\
capability & 363 & \textbf{17.3\%} & 510 & \textbf{24.3\%} \\
\bottomrule
\end{tabular}
\caption{The 2{,}100 model-attributed agent-mode losses under both admissible orderings of the two overlapping
definitions in \S\ref{sec:classdef}. Under the ordering we adopt, capability is the smallest class; under the
alternative it is the second largest and verification becomes the smallest. Transport is the largest class
either way, and that is the finding that does not depend on the choice.}
\label{tab:taxonomy}
\end{table}

Table~\ref{tab:taxonomy} is the paper's central result and we state its limits with it. The capability-threshold
framing predicts that the capability class should dominate. Under our adopted ordering it is the smallest of
four, and more than four losses in five come from runs that engaged the tools. Under the alternative ordering
capability rises to 24.3\% and the claim weakens to a different one: transport alone still exceeds it, and the
majority of losses still come from runs that used the tools, but capability is no longer the smallest class.
Readers who prefer the stricter reading should take the second pair of columns and the weaker claim.

\textbf{Why this table covers agent mode only.} Planning is toolless by construction, so every one of its 986
runs invokes zero tools and every one of its 182 losses lands in clock or capability by definition, and none can
ever enter transport. Including planning would therefore let an entire surface inflate the two classes our
headline compares while being structurally barred from the third. Reporting agent mode alone removes that
artefact, and it moves the result in the direction that makes it harder for us, not easier: it raises the class
we are arguing for and lowers the one we are arguing against only by removing runs that could never have
contradicted us. For completeness, over all 3{,}142 losses including planning and unattributed runs the adopted
ordering gives transport 33.3\%, clock 26.5\%, verification 20.6\% and capability 19.7\%.

The transport class is not marginal engagement. Across those 761 runs the median tool-invocation count is 2 and
the total is 1{,}926 invocations against 1{,}219 further declined calls: these are runs that inspected schemas,
ran reads and examined plans, and then failed to file.

\subsection{How a cell is scored, and how much the choice is worth}
\label{sec:scoring}

A cell's history is not one experiment. It spans code changes and configuration changes, so how a scorer searches
that history for five consecutive passes decides the answer. The released scorer implements three modes and we
report the gap between them, because it is large and because the obvious implementation is the one that
overstates.

\emph{Pooled} scans a cell's entire run history for any streak of five consecutive passes. It is the obvious
implementation. \emph{Session} splits the history into sittings and requires the streak to fall inside one, but
allows different cells to come from different sittings. \emph{Row}, which is our protocol, splits the history into
sittings, scores each sitting on its own and takes the best, so all six cells must lock inside one sitting.

On this corpus the difference is not marginal. Two models score 6/6 and 30/30 pooled and 23/30 under row. When
those two were re-read by hand, six surfaces in one sitting, they returned 24/30 and 21/30, which is what row
predicts and not what pooled does. A pooled streak says a model passed five times running under \emph{some}
mixture of conditions, which is not a claim a deployment can act on. Every per-model figure in this paper is a
row score, and the released scorer reproduces all three so that a reader can see the gap rather than take our
word for it.

\subsection{Per-model results}

\begin{table}[htbp]
\centering\scriptsize
\begin{tabular}{@{}lrrr@{}}
\toprule
Model & Runs & Cells passed (of 30) & Surfaces locked (of 6) \\
\midrule
\code{qwen3:4b} & 73 & 30/30 & 6/6 \\
\code{qwen3.6:35b} & 137 & 30/30 & 6/6 \\
\code{qwen3.5:27b} & 30 & 30/30 & 6/6 \\
\code{qwen3-coder:30b} & 71 & 30/30 & 6/6 \\
\code{gpt-oss:20b} & 272 & 30/30 & 6/6 \\
\code{granite4.2:3b} & 131 & 28/30 & 5/6 \\
\code{gemini-3.5-flash-lite} & 90 & 28/30 & 5/6 \\
\code{qwq:32b} & 50 & 28/30 & 4/6 \\
\code{deepseek-r1:8b} & 383 & 27/30 & 5/6 \\
\code{mistral-small3.2:24b} & 548 & 25/30 & 5/6 \\
\code{granite4:3b} & 85 & 25/30 & 5/6 \\
\code{magistral:24b} & 49 & 25/30 & 4/6 \\
\code{glm-4.7-flash:latest} & 217 & 25/30 & 4/6 \\
\code{mistral-nemo:12b} & 398 & 24/30 & 4/6 \\
\code{llama3.1:8b} & 629 & 23/30 & 4/6 \\
\code{granite4.1:3b} & 70 & 23/30 & 4/6 \\
\code{cogito:8b} & 551 & 23/30 & 3/6 \\
\code{ministral-3:3b} & 119 & 23/30 & 2/6 \\
\code{phi4-mini:3.8b} & 30 & 22/30 & 4/6 \\
\code{command-r7b:7b} & 115 & 20/30 & 3/6 \\
\code{lfm2:24b} & 45 & 17/30 & 3/6 \\
\code{hermes3:8b} & 145 & 17/30 & 3/6 \\
\code{glm4:latest} & 223 & 17/30 & 3/6 \\
\code{qwen3.5:2b} & 60 & 16/30 & 2/6 \\
\code{qwen3-vl:8b} & 81 & 14/30 & 2/6 \\
\code{qwen3:1.7b} & 55 & 12/30 & 1/6 \\
\code{deepseek-r1:14b} & 30 & 12/30 & 1/6 \\
\code{mistral-small3.1:24b} & 35 & 10/30 & 2/6 \\
\code{qwen2:7b} & 33 & 8/30 & 1/6 \\
\code{mistral:7b} & 30 & 5/30 & 1/6 \\
\code{qwen3-1.7b} & 30 & 0/30 & 0/6 \\
\bottomrule
\end{tabular}
\caption{Every model with a complete six-surface reading in the released corpus, scored by the released scorer.
A surface is locked only by five consecutive passes, scored in row mode (\S\ref{sec:scoring}) over each
model's whole history. Nine of the 31 lock every surface; one locks none. This table is the denominator: it
includes the models that did not work, which a roster of supported models by construction cannot. It is not
comparable cell by cell with the intervention table, which reports one specific 30-run sitting rather than a
model's best sitting, and the two can therefore differ by a cell or two for the same model.}
\label{tab:leaderboard}
\end{table}

\subsection{Where the refusals are}

\begin{table}[htbp]
\centering\small
\begin{tabular}{@{}lr@{}}
\toprule
Tool and reason & Count \\
\midrule
\code{compose\_report : INVALID\_TOOL\_INPUT} & 6{,}243 \\
\code{compose\_report : UNVERIFIABLE\_EVIDENCE} & 3{,}464 \\
\code{tool : database-error} & 1{,}110 \\
\code{present\_answer : INVALID\_TOOL\_INPUT} & 1{,}032 \\
\code{present\_answer : ANSWER\_NOT\_A\_DATA\_READ} & 696 \\
\code{recommend\_change : INVALID\_TOOL\_INPUT} & 613 \\
\code{recommend\_change : RECOMMENDATION\_SHAPE\_MISMATCH} & 248 \\
\code{compare\_plans : INVALID\_TOOL\_INPUT} & 152 \\
\code{compare\_plans : UNVERIFIABLE\_PLAN} & 139 \\
\midrule
All other reasons & 311 \\
\textbf{Total} & \textbf{14{,}008} \\
\bottomrule
\end{tabular}
\caption{Refused calls by tool and reason. \code{compose\_report} alone accounts for 9{,}707 refusals, 69.3\% of
the total, split between calls whose \emph{shape} did not validate and calls whose \emph{citations} did not
resolve. The single tool that turns completed work into a delivered answer is where two thirds of all refusals
occur.}
\end{table}

\section{What transport failures actually look like}
\label{sec:transport}

The taxonomy says the largest class is work done and delivery failed. This section reports what the model text
and arguments show inside it, because the mechanism is specific and repairable.

\subsection{The call written into the wrong channel}

Of 69 stopping turns whose text we examined, 27 contained a complete tool call for a tool the run held, 18 of
them the reporting tool whose absence \emph{is} the missing-report verdict. A representative case, verbatim from
the ledger:

\begin{lstlisting}
{"action": "compose_report", "arguments": {"claims": [
  {"claim": "The current query plan uses a full table scan, which can be slow for large tables.",
   "evidence": [{"source": "artifact", "correlationId": "935c1381-d06a-4692-a02a-fd466b7ad62a"}]},
  {"claim": "Adding an index on first_name and last_name will improve performance ...",
   "evidence": [{"source": "artifact", "correlationId": "935c1381-d06a-4692-a02a-fd466b7ad62a"}]}
}]}
\end{lstlisting}

The claims are true, the evidence arrays are populated, and the correlation id is one the run genuinely produced.
The model wrote this into the assistant text channel instead of calling with it, and the run ended having done
every part of the work except the transport.

None of the 27 parsed as JSON. Twenty-four failed identically, on an array whose last member closes twice.
Twenty used the key \code{name} and seven used \code{action}. Any recovery path that requires a JSON parse to
succeed sees none of them.

\subsection{Displaced fields}

Capturing the arguments of refused calls on two models over the two weakest surfaces produced 17 refusals in five
runs per cell. Nine returned the same three-field refusal message. Four of those nine, all from one model, were the same
displaced shape, differing only in the correlation id they cited:

\begin{lstlisting}
{"change": "CREATE INDEX idx_employee_emp_no ON employee(emp_no);",
 "reason": "Creating an index on the employee table will improve the query performance ...",
 "evidence": {"correlationId": "772210b6-dcd8-4423-ae2b-21fd49a4b600", "source": "artifact"}}
\end{lstlisting}

The declared schema is

\begin{center}
\code{\{change: "index"|"rewrite", statement: string, rationale: string, evidence: Evidence[]\}}
\end{center}

\noindent In those four every value is correct and three are displaced, each in its only plausible
direction: the statement sits in the field named \code{change}, the rationale is called \code{reason}, and one
evidence item arrived as itself rather than as a list of one. The model was told three fields were missing,
about a call carrying all three values, four times. The remaining five of the nine came from a second model in
three further shapes, \code{\{column\_name, index\_name, table\_name\}}, \code{\{column\_name,
index\_name\}} and \code{\{sql\}}, carrying between one and three of the required values against the same
refusal text. A further five refusals carried a correct artifact id under a key one affix off. Across all
released captures exactly one pair of calls is byte-identical, so the repetition here is of a shape rather than
of a string.

\subsection{The refusal that instructs the model to discard its answer}
\label{sec:remove}

The clearest defect we found, and the one with the largest single-cell effect. On one 3B model's data-analysis
cell, 0/5 with every loss recording a filed report and no presented answer while its other five surfaces locked,
the captured arguments show a run that had both halves of the answer and filed one on the wrong tool:

\begin{lstlisting}
present_answer  <-  {"artifact": "bfe7ca93-2bac-4df8-a091-a3d21bfa0454"}
compose_report  <-  {"claims": [...], "presentation": {"kind": "table"}}
\end{lstlisting}

The capture holds nine records for this cell in three identical groups: two calls to the answering tool carrying
only the artifact and refused for a missing presentation, then one call to the reporting tool \emph{carrying} a
presentation and refused with \code{the arguments object: remove presentation}. Put together, any such pair is
complete and correct.

The contradiction is structural and does not depend on what the model did next: the server demands
\code{presentation} on one tool, forbids it on the sibling tool, and scores the run as having no answer when the
field is absent, while the model has in fact composed one. \code{remove} is the correct word for a key that
belongs nowhere and the wrong word for a key that is a field of a neighbouring tool, and the run is failed for
the absence either way.

We state explicitly what the released capture does \emph{not} establish. It carries no run identifier, no turn
ordinal and no timestamp, so the order of these nine records cannot be recovered from it, and in file order the
presentation-less calls precede the remove instruction rather than follow it. We therefore make no claim that
the model deleted the field in response to being told to. The defect is the contradictory contract, which is
visible without the ordering.

\section{Interventions and measured effect}
\label{sec:interventions}

Five changes were made to the server on the strength of the above. None touches a model, a prompt, a temperature
or a task.

\begin{table}[htbp]
\centering\small
\begin{tabular}{@{}p{0.3cm}p{5.3cm}p{7.6cm}@{}}
\toprule
\# & Defect & Change \\
\midrule
1 & Text-channel calls invisible unless strictly parseable JSON keyed \code{name} & Recognise the call by its naming rather than by parsing the document \\
2 & SQL placed in the enum field \code{change} & Read the kind from the statement; prose is still refused \\
3 & \code{rationale} sent as \code{reason}; evidence sent as one object or as a JSON string & Exactly-one-absent and exactly-one-surplus is an unambiguous rename; a lone object is a one-item list \\
4 & A near-miss key named in the refusal for months and never applied & Apply the rename on flat two-field calls; leave nested calls to be told \\
5 & \code{remove <field>} told models to discard a sibling tool's field & \code{remove} reserved for keys that belong nowhere; a field owned by another tool is answered with where it goes \\
\bottomrule
\end{tabular}
\caption{The five server-side interventions. See \S\ref{sec:remove} for the fifth.}
\end{table}

\begin{table}[htbp]
\centering\small
\begin{tabular}{@{}lrr@{}}
\toprule
Model & Before & After \\
\midrule
\code{granite4.2:3b} & 21/30 & \textbf{28/30} \\
\code{mistral-nemo:12b} & 18/30 & \textbf{26/30} \\
\code{llama3.1:8b} & 10/30 & \textbf{24/30} \\
\code{cogito:8b} & 0/30 & \textbf{21/30} \\
\code{glm4:latest} & 0/30 & \textbf{16/30} \\
\code{deepseek-r1:8b} & 18/28 & \textbf{24/28} \\
\bottomrule
\end{tabular}
\caption{Cells open before the change, re-read after, same configuration and objectives. Scores are runs whose
verdict was \code{answered}, resolved from the sweep logs' run identifiers against the released dataset; the logs
themselves record only process exit, which is not a pass. Two of \code{deepseek-r1:8b}'s thirty scheduled runs
never started, so its denominator is 28 and two of its cells are short of the five runs a lock requires. The
corpus-wide attrition is 89 scheduled runs that never started.}
\end{table}

We report two qualifications that cut against these figures. First, the widened text-channel reader,
intervention 1, measured in isolation on four cells produced $+1$, $0$, $0$, $-1$, which is inside variance. It
is a correct fix serving a real 27-run population and it is not what moved the table; the argument-shape readers
are. Second, part of the movement for three of these models is not attributable to code at all: their prior
figures were unions of readings taken across days and settings, which \S4.1 explains is not a valid row.
Separating those two effects cleanly would require re-reading every prior row in one sitting, which we have not
done. These are therefore field observations with known confounds, not an ablation.

\section{Threats to validity}

\subsection{Unbounded serving context is a memory confound, and we hit it}
\label{sec:memory}

With no context length configured, Ollama admits each model at its full advertised context. One 12B model, 7.1\,GB
on disk, was resident at 51\,GB with a 262{,}144-token context on a 64\,GB machine, and free memory fell to 6\%.

Runs taken in that state are not measurements. A 3B model's confirmation pass read its planning cell 0/5, all
five runs hitting the 90-second turn limit having invoked no tool, a row that reads in any log exactly like a
model that cannot plan. Capped at 32{,}768 the same model is 5.1\,GB and the same cell reads 5/5.

Two properties make this dangerous for published work. System-wide free memory reads healthy, 76\% in one
sample, while a single model holds most of RAM, so the usual guard misses it. And the desktop application
restarts its own server and reclaims the port, so a capped server fails to bind, logs an address-in-use error,
and the uncapped server answers. That looks like success. The serving engine's own process listing, with its
context column, is the only check we found that does not lie.

All figures in Section~\ref{sec:results} were taken before the cap was applied. The per-model and per-surface
rates should therefore be read as a lower bound with unknown per-model bias, since large-context models were
penalised more than small ones. Re-reading the corpus under a fixed cap is the first item of future work. We
expect the taxonomy shape to survive it, because the class it should shrink is the clock class, and under our
adopted ordering the clock class is not the largest. We note that this expectation cuts against us under the
alternative ordering in Table~\ref{tab:taxonomy}, where shrinking the clock class moves runs into capability.

\subsection{A related confound: swap}

Separately, one sweep read a 24B model's analysis cell at 0/5 with runs of 244 to 355 seconds against 41 to 48
seconds for the same cell earlier. Free memory read 53\% throughout while swap sat at 19\,GB of 20\,GB with 1.45
million pageouts. Watchdogs that check free memory do not see this. What works is watching swap growth: a full
but static swap file costs nothing, whereas swap being written while a model runs is the machine choosing
between the model and everything else.

\subsection{Single machine, single database, single harness}

One machine, one schema, one agent implementation. The surface ordering and the taxonomy are properties of
\emph{this} tool contract. A contract with fewer structured artifacts per surface should show a smaller transport
class, which is a testable prediction rather than a caveat.

\subsection{Unequal sampling and unattributed runs}

Run counts per model range from 3 to 629 and were allocated by operational interest rather than by design:
models that looked close to completion were run more. Per-model rates therefore carry very unequal confidence and
the corpus is not a randomised comparison. The taxonomy shares, computed over 2{,}100 agent-mode losses, are the more
robust figure. Separately, 3{,}248 runs predate the ledger event that records the model and carry no attribution; they
enter only the whole-corpus figures. Two model tags in the corpus differ only in punctuation and may denote the
same model, which would make the named count 39 rather than 40.

\subsection{The run deadline is a hard boundary we did not move}

One 8B model's query-optimization cell passes at 443 and 444 seconds and fails at the 450-second run deadline.
Raising the deadline would likely close it. We did not, because the deadline is part of the shipped product
configuration, and a model that only passes outside it has not been shown to work for a user. This is a
deliberate choice that costs us cells.

\subsection{Uncertainty, and what it does not cover}

These are uncontrolled field observations. Runs were not randomised, run counts per model were allocated by
operational interest, and the corpus spans five code checkouts, so nothing here identifies a cause. Sampling
uncertainty can still be quantified, and it should be, because the naive figure misleads in a specific direction.

Treating the 2{,}100 losses as independent gives narrow intervals: transport 36.2\% [34.2, 38.3], clock 25.8\%
[24.0, 27.7], verification 20.7\% [19.0, 22.5], capability 17.3\% [15.7, 19.0], all Wilson 95\%. They are
independent only if failure mode is unrelated to model, which it plainly is not. Resampling the 33 contributing
models rather than the runs, which is the correct unit, widens them to transport [22.0, 49.1], clock [10.1,
44.6], verification [8.5, 32.9] and capability [6.6, 31.4], and transport is the largest class in 74.5\% of
resamples rather than in all of them. Across the 22 models contributing at least twenty losses, transport is the
largest class for 10 and capability the smallest for 13.

We therefore state the class ordering as a property of this corpus and not as a finding that generalises across
models. What survives the clustered treatment is the engagement result: 1{,}590 of 2{,}100 losses, 75.7\% [73.8,
77.5], came from runs that had invoked at least one tool; the majority holds in 99.7\% of clustered resamples and
in 15 of the 22 models with at least twenty losses. That is the claim the paper rests on, and it is the one that
contradicts the capability-threshold framing.

The per-model before and after readings rest on thirty runs each, six cells of five rather than thirty
independent trials. A sign test on six of six models improving gives $p = 0.031$. The cells were selected for
re-reading because they were open, that is, because they had scored low, so regression to the mean predicts a
positive mean delta even under a zero true effect; the two rows that began at 0/30 are the ones a floor protects
from that. We report the six as field observations, not as an estimate of an effect size.

\section{Discussion}

\paragraph{Which mechanism?}
Our first reading of the transport class was informational: refusals failed because they stated the expectation
and not what had arrived, so there was nothing in them to act on. The displaced-field data in
\S\ref{sec:transport} does not support that reading. In those runs the refusal named the missing fields
explicitly and one model resent the same displaced shape four times while a second answered the same refusal text
with four further shapes. A message containing the information needed
to repair the call was not sufficient. That is what Gumaan's counterfactuals predict: if most of the damage is
the presence of the failed call's surface form in the context, then improving the accompanying sentence addresses
the smaller term. It is also consistent with what actually moved our numbers. Interventions 2 to 5 did not
produce better sentences; three of them changed what the server \emph{accepts}, and the fifth replaced an
instruction that was actively wrong. We read our data as evidence against a purely informational account.

\paragraph{Against the capability attribution.}
AgentBench attributes the open-versus-commercial gap to reasoning and instruction following. Our results do not
refute that, and our own corpus contains models that never became usable. What they suggest is that any such
attribution is only as good as the harness it was measured through, and that a benchmark which fixes its
rejection messages and its argument parsing as part of the apparatus is measuring the pair rather than the model.
Redesigning tool error feedback is already a standing recommendation~\cite{xiong2025butterfly}, and
ToolMisuseBench~\cite{sigdel2026toolmisuse} already treats the feedback returned after a failed call as a
declared condition; what we add is a measurement, in a deployed system, of what making that change is worth. We
would extend the practice to the leaderboards that report per-model results, so that the rejection-message text
and the parser's tolerance are published as benchmark artefacts in their own right.

\paragraph{Practical advice.}
Record the arguments of refused calls somewhere, even when they must not enter the audit ledger: a shape that
cannot be seen cannot be fixed. Separate refused-at-the-transport from tried-and-failed before reporting a zero.
Check the serving engine's resident context before trusting a timeout. And read a model's cells in one sitting,
because a union of readings is not a result.

\section{Conclusion}

The question of whether a model is big enough to drive an agent is the wrong first question. In 8{,}199 runs,
under the classification we adopt and argue for, only one loss in six is a model that failed to engage; under the
stricter alternative it is closer to one in four. Either way the majority of losses are models that engaged,
established something true, and lost it to a field name, a brace, a channel, a clock, or a refusal that told them
to throw the answer away.

That is an interface problem, and interface problems have the useful property that fixing them once helps every
model at once. Five server-side changes, none model-specific, moved six models by 6 to 21 cells out of 30 on the
same hardware, with the same prompts, on the same tasks. The distance left is measured in cells, not in
parameters.

\section*{Author contributions (CRediT)}
\noindent
\textbf{Cevheri Bozo\u{g}lan}: conceptualisation, software, methodology, supervision, writing.
\textbf{Yusuf G\"{u}ndo\u{g}du}: software, methodology, investigation, formal analysis, data curation, writing.
\textbf{Abdullah Kaya}: investigation, data curation.
\textbf{Koray \c{S}irin}: software, validation.

\section*{Data and code availability}
The application and its agent are at \url{https://github.com/libredb/libredb-studio} under the MIT licence.
The measurement artefacts are \emph{not} in that repository: the raw run ledgers are written to a working
directory the project gitignores, so they had never been published before this paper.

The complete measurement record is released as a dataset: 8{,}199 run records, 110{,}711 ledger events and
14{,}008 refusal records, with the exporter that produces them from the raw ledgers, a scorer that reproduces the
per-model table, and a verification script that regenerates every figure in this paper and exits non-zero if any
disagrees. It is published as a dataset~\cite{libredb2026dataset},
\url{https://doi.org/10.57967/hf/10485}, because the event stream alone exceeds the size a paper submission
can carry. A reduced bundle accompanies this submission directly: the verification and scoring
scripts, the captured arguments of refused calls, and the 160 per-cell sweep logs.

We ask readers to run the verifier before trusting a number here. Every figure in this paper was produced by it,
and the classification in \S\ref{sec:classdef} agrees with the \code{loss\_class} field of the released dataset
on all 2{,}194 model-attributed losses, of which 2{,}100 are agent-mode.

\section*{Licence}
This paper is released under CC BY 4.0. The dataset, the exporter, the scorer and the verifier are released under
the MIT licence, as is the application.

\section*{Conflict of interest}
All authors are contributors to LibreDB Studio, the software evaluated. No external funding supported this work.

\end{document}